\documentclass[aip,pof,reprint]{revtex4-1} 
\usepackage{graphicx} 
\usepackage[usenames]{color}
\usepackage{amsmath,amssymb}
\usepackage{gensymb}
\begin{document} 

\title{Origin of line tension for a Lennard-Jones nanodroplet}

\author{Joost H. Weijs$^1$, Antonin Marchand$^2$, Bruno Andreotti$^2$, Detlef Lohse$^1$, and Jacco H. Snoeijer$^1$}
 
\affiliation{
$^{1}$Physics of Fluids Group and J. M. Burgers Centre for Fluid Dynamics,
University of Twente, P.O. Box 217, 7500 AE Enschede, The Netherlands\\
$^{2}$Physique et M\'ecanique des Milieux H\'et\'erog\`enes, UMR
7636 ESPCI -CNRS, Univ. Paris-Diderot, 10 rue Vauquelin, 75005, Paris,
France}

\date{\today} 

\begin{abstract}
The existence and origin of line tension has remained controversial in literature. To address this issue we compute the shape of Lennard-Jones nanodrops using molecular dynamics and compare them to density functional theory in the approximation of the sharp kink interface. We show that the deviation from Young's law is very small and would correspond to a typical line tension length scale (defined as line tension divided by surface tension) similar to the molecular size and decreasing with Young's angle. We propose an alternative interpretation based on the geometry of the interface at the molecular scale.
\end{abstract} 

\maketitle

\section{Introduction}
The development of microfluidics in the last decade has renewed the interest for a thermodynamical concept introduced by Gibbs in his pioneering article: line tension~\cite{Gibbs}. By analogy with surface tension, which is by definition the excess free energy per unit surface of an interface separating two phases, line tension is the excess free energy per unit length of a contact line where three distinct phases coexist. The variation of a system free energy $F$  therefore presents three contributions, a bulk contribution when the volume $V$ is varied, a surface contribution when any interface area $S_i$ is varied and a line contribution when the contact line length $L$ is varied:
\begin{equation}
\label{eq:dF}
dF=PdV+\sum_i \gamma_i dS_i+\tau dL \; .
\end{equation}
Here, we use a summation to indicate that one has to take all interfaces into account (liquid-solid, liquid-vapour, and solid-vapour). The stability of deformable surfaces, such as a liquid-vapor or liquid-liquid interface, necessarily requires a positive surface tension. Although the shape of the contact line is deformable as well, the line tension cannot be inferred from a stability argument~\cite{Guzzardi}. In addition, there are conceptual problems defining line tension properly~\cite{SchimmeleEPJE,SchimmeleJCP}.

\begin{figure}[tbp]
\begin{center}
\includegraphics{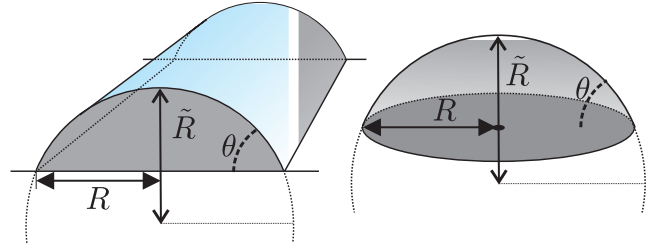}
\caption{\label{fig:schem}Schematic of two drops of same size, one cylindrical cap shaped (left) and the other spherical cap shaped (right). For small volumes, the contact angle $\theta$ for the spherical cap is affected by line tension, while $\theta$ is constant for the cylindrical cap.}
\end{center}
\end{figure}

The simplest system in which a line tension effect may be observed is a liquid drop on a solid substrate, in partial wetting conditions.  Consider a drop whose shape is a spherical cap characterised by its contact line radius $R$ --~seen from the top~-- and its contact angle $\theta$. The drop volume is $V=\frac{1}{3}\pi \tilde{R}^3 \left( 2-3\cos\theta + \cos^3\theta\right)$, the liquid-vapour area $S_{LV}=2\pi \tilde{R}^2(1-\cos \theta )$, the solid-liquid area $S_{SL}=\pi R^2$ and the contact line length $L=2\pi R$. Here we defined $\tilde{R}$ as the radius of curvature of the spherical cap: $\tilde{R}=R / \sin \theta$, see also Fig.~\ref{fig:schem}. When minimizing the free energy with respect to $\theta$ at constant volume ($PdV=0$), one gets \cite{Pethica}:
\begin{equation}
\cos \theta=\frac{\gamma_{SV}-\gamma_{SL}}{\gamma}-\frac{\tau/\gamma}{R}=\cos\theta_{Y}-\frac{\tau/\gamma}{R}\;,
\label{eq:modyoung}
\end{equation}
where $\theta_Y$ is the Young contact angle and $\gamma_{SL}$, $\gamma_{SV}$, $\gamma$ the solid-liquid, solid-vapour, and liquid-vapour surface tension, respectively.  Note that \eqref{eq:modyoung} only holds for spherical cap-shaped droplets -- the contact line of a cylinder-shaped drop has zero curvature, which means that the contact angle $\theta$ is unaffected by line tension and independent of drop size. In this derivation, we did not take any interface curvature effects into account (such as Tolman-corrections on $\gamma$), if these become comparable in magnitude to line tension the measured $\tau$ from eq.~\ref{eq:modyoung} cannot be considered `pure' line tension, but rather an apparent line tension~\cite{SchimmeleEPJE,SchimmeleJCP}. From \eqref{eq:modyoung} one can see that when $\tau$ is positive, drops will present a larger contact angle than Young's angle.

Theoretical predictions on the strength of line tension are based on calculating the free energy (per unit length) associated with the contact line, using statistical mechanics (e.g. using density functional theory \cite{GettaDietrich,BauerDietrich}) or a model based on interface displacement \cite{Dobbs/Indekeu,Churaev}. These analyses predict the value of line tension to be in the range $10^{-12}$ to $10^{-10}$~J/m. Of particular interest is the behaviour near the wetting transition ($\theta\rightarrow 0$), for which $\tau$ can vanish or diverge depending on the details of the interaction~\cite{DeGennes,Amirfazli,Indekeu,Indekeu1994}.

A large amount of experimental work has been done to determine the magnitude of line tension. The most direct way is to measure the contact angle as a function of contact line curvature, and thus droplet size~\cite{DrelichMiller,DrelichMiller92,Gaydos,Li1990,Vera-Graziano}. Using the modified Young-equation from \eqref{eq:modyoung} $\tau$ can then be calculated. Due to the small length scales involved for the measurement of $\tau$, the observed values for $\tau$ vary greatly in magnitude: both negative and positive values as low as $10^{-11}$~J/m and as high as $10^{-5}$~J/m have been reported. The reason for the huge variation is that determining the contact angle is notoriously difficult due to contact angle hysteresis caused by surface inhomogeneities~\cite{Drelich}. The slightest amount of surface inhomogeneities can cause a severe overestimation of $\tau$. Indeed, contaminated surfaces can even lead to an apparent change of sign of $\tau$~\cite{Checco,Li}. Historically, droplets were used to measure line tension. Recent developments on surface nanobubbles allowed to detect a similar size dependence of the contact-angle of the nanobubbles~\cite{Borkent,Yang,Kameda,Nakabayashi}.
Because of the difficulty of exact contact angle measurements at the required scale (1-100 nm), alternative methods have been developed, for example by calculating the effective potential near the contact line by measuring the deviation of the liquid surface from a wedge shape~\cite{Pompe,Zorin}. For reviews on experimental methods and results see refs.~\cite{Gaydos,Amirfazli}.

In this paper we adopt the usual experimental method to determine the line tension, by measuring $\theta$ against $1/R$, in a theoretical setting. We will perform this measurement for different equilibrium contact angles $\theta_{Y}$, to study the dependence of the line \emph{tension length} ($\ell$) on $\theta_{Y}$. The tension length is defined as:
\begin{equation}
\label{eq:ell}
\ell\equiv -\frac{\tau}{\gamma}\;.
\end{equation}
Since we always find negative values of $\tau$ for the Lennard-Jones drops studied in this paper, the minus sign is added to ensure that the tension length is always positive. 

We perform these measurements for both 3D (spherical cap-shaped) and 2D (cylindrical-shaped) droplets, to compare similar sized droplets with and without contact line curvature. In the first part of this paper, we investigate line tension by means of molecular dynamics simulations of a Lennard-Jones droplet, which has the added advantage that no assumptions have to be made as is required for most analytical approaches. In this sense, these simulations are like an experiment, but with unprecedented accuracy and without surface inhomogeneity. Since the expected tension length is of the order of the molecular size, the problem also has the right scale for molecular dynamics. Line tension has been observed in molecular dynamics studies before~\cite{Halverson,Ingebrigtsen}, but a systematic study has to our knowledge not been carried out. In the second part of the paper, we analyse the existence and origin of line tension using the density functional theory (DFT) in the approximation of the sharp kink interface. Finally, we will calculate the line tension using a geometric interpretation based on missing bonds in a wedge-shaped interface. We show that the deviation from Young's law is very small and would correspond to a line tension of a fraction of the molecular size.
 
\section{Nanodrops from Molecular Dynamics}\label{sec:MD}

\subsection{Numerical setup}
\begin{figure}[t!]
\begin{center}
\includegraphics{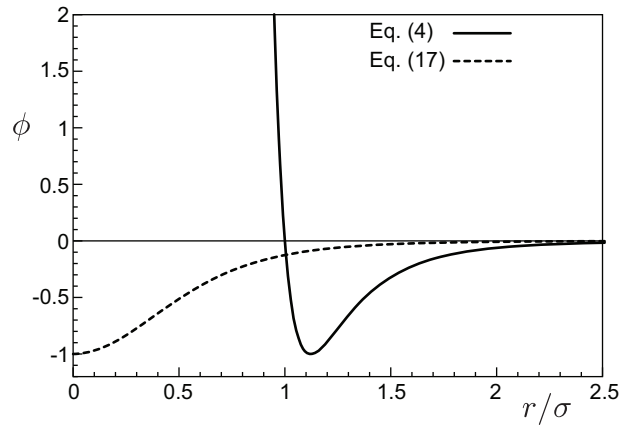}
\caption{Interaction potentials $\phi(r)$ for the Lennard-Jones particles \eqref{eq:LJ} (solid line) and the potential $\rho^{(1)}\rho^{(2)} g_r(r)\phi(r)$ used for the DFT-calculations \eqref{eq:regpot} (dashed). Note that the DFT potential is regularized to account for vanishing $g_r(r)$ when $r\rightarrow 0$.}
\label{fig:pots}
\end{center}
\end{figure}
We perform Molecular Dynamics (MD) simulations on nanodrops, using the \textsc{Gromacs} software package~\cite{gromacs}. We simulate binary systems in which two types of particles exist: fluid particles that can move around either in the gas or liquid phase, and solid particles which are frozen on an fcc lattice and constitute the solid substrate,Fig.~\ref{DropsMD}(a). The simulations are done in the NVT-ensemble, where the temperature is held at 300K using a thermostat, which is below the critical point for a Lennard-Jones fluid with the interaction strengths used. All particle interactions are defined by the Lennard-Jones (LJ) potential:
\begin{equation}
\label{eq:LJ}
\phi^{LJ}_{ij}(r)=4\epsilon_{ij} \left[ \left( \frac{\sigma_{ij}}{r} \right)^{12} - \left(\frac{\sigma_{ij}}{r}\right) ^6 \right] \;,
\end{equation}
see also Fig.~\ref{fig:pots}. Here, $\epsilon_{ij}$ is the interaction strength between particles $i$ and $j$ and $\sigma_{ij}$ the characteristic size of the molecules. This size is chosen the same for all interactions, $\sigma_{ij}=\sigma$. The potential function is truncated at a relatively large radius ($r_c = 5\sigma$) where $\phi^{LJ}$ is practically zero. The timestep is chosen at $dt=\sigma\sqrt{m/\epsilon_{ll}}/200$, with $m$ the mass of the particles. The fluid particles are initially positioned on an fcc-lattice near the substrate, but are free to move around and relax towards an equilibrium droplet shape (Fig.~\ref{DropsMD}). Periodic boundary conditions are present in all directions. To study the effect of line tension we consider two different systems. In the "3D" case the dimensions of the system are chosen large enough to ensure that the droplet does not interact with itself, resulting in an isolated droplet with the shape of a spherical cap. In the "2D" case the system size in the $x$-direction (parallel to the substrate) is only 15$\sigma$ leading to an infinitely long cylindrical-cap shaped droplet. The small length is required to suppress the Rayleigh-instability, which is only effective at wavelengths $\lambda > 2\pi R$. Now, as there is no contact line curvature, line tension has no effect.

The wettability of the substrate (and thus, the equilibrium contact angle) is tuned through the parameter ratio $\epsilon_{LS}/\epsilon_{LL}$, where the subscript indicator $S$ denotes the solid (fixed) particles and $L$ the liquid particles. A higher value for $\epsilon_{LS}/\epsilon_{LL}$ results in a large attraction of fluid particles to the substrate, and thus a more wetting substrate. A range of contact angles can be explored in this way, as shown in table I. The obtained contact angles and compare well to those found in previous studies~\cite{Barrat,Dammer}. Depending on the size of the droplet, the effect of layering inside the liquid (Fig.~\ref{DropsMD}) limits the range where reliable contact angle measurements can be performed. In practice, this limits the analysis to contact angles larger than approximately $70^\circ$. 

\subsection{Cylindrical vs spherical caps}
\begin{figure}[htbp]
\begin{center}
\includegraphics{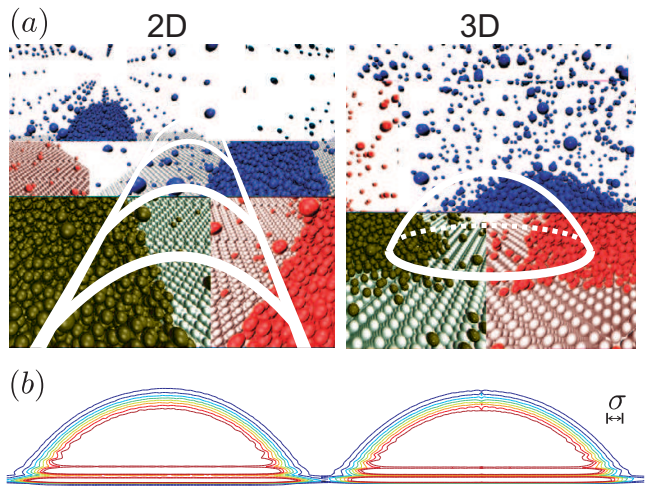}
\caption{(a) Snapshots from Molecular Dynamics simulations of a 2D, cylindrical cap-shaped droplet (left), and a 3D, spherical cap-shaped droplet (right). The light spheres represent the immobilized solid particles, forming the substrate to which the droplet attaches. The darker spheres represent the mobile fluid particles. The lines are a guide to the eye. Several periods of the 2D droplet are shown (periodic boundary conditions), causing the same particle to be printed multiple times. These drops were simulated using identical interaction parameters ($\frac{\epsilon_{SL}}{\epsilon_{LL}}=\frac{2}{3}\Rightarrow\theta_Y\approx 65^\circ$), and differ in shape only because of the difference in the periodic boundary conditions. (b) Isodensity contours measured using statistical averaging from the droplets shown in the top row. The contact angle and overall shape of the two drops is almost identical, requiring a precise measurement to observe the effect of line tension.}
\label{DropsMD}
\end{center}
\end{figure}
Fig.~\ref{DropsMD}(a) shows the shape of two nano-drops ($\theta_Y=65^\circ$) with similar radii $R$ (as seen from the top), but with a different geometry. The cylindrical droplet on the left is formed in the quasi-2D system (several periods shown). The spherical cap shaped droplet on the right is simulated in a fully 3D system. Fig.~\ref{DropsMD}(b) shows the isodensity contours from the same droplets. One observes that these cross-sectional shapes are already very similar, indicating that line tension is indeed a weak effect, even for such nanodrops. We will now perform careful and precise contact angle measurements in order to quantify line tension. 

\subsection{Measurement of contact angle}
\begin{figure*}[htbp]
\begin{center}
\includegraphics{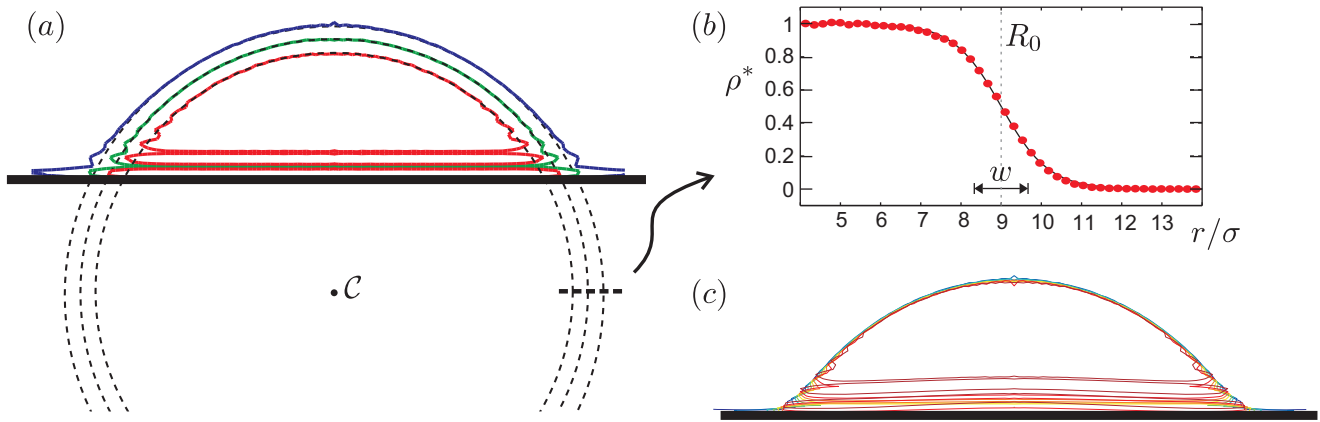}
\caption{Isodensity contours of a Lennard-Jones droplet. (a) A selection of isodensity contours (same drop as in Fig.~\ref{DropsMD}, $R_0\approx 9\sigma$, $\rho^\ast=0.3,0.5,0.7$) fitted by circles (dashed). The cirles turn out to be concentric which we use to collapse the contours into one single shape. (b) Density profile \eqref{eq:dens} fitted to measured $\rho^\ast(r)$, where we can see that the interface is several molecular diameters thick. With this fit we can shift all contours by using \eqref{eq:dens}. (c) Rescaled iso-contours nicely collapse on a single curve, which allows us to define the interface in a precise way.}
\label{fig:collapse}
\end{center}
\end{figure*}
To perform precise contact angle measurements we first compute the density field by averaging over time and over space (translational or rotational symmetry). During this averaging, we compensate for any center of mass motion of the droplet parallel to the substrate by moving the droplet such that the center of mass is stationary throughout the averaging procedure. When the droplet has reached its equilibrium state (Fig.~\ref{DropsMD}), the density profiles are calculated by time-averaging over 1,000,000-10,000,000 time steps until the density field has converged. Using real world parameters for Argon as the fluid this would correspond to 2 - 20 nanoseconds. This leads to droplet shapes as shown in the bottom row of Fig.~\ref{DropsMD}. The part of the droplet that is close to the substrate is subject to layering~\cite{Dammer}: the density oscillates as a function of height. To avoid interference from this effect, we ignore this part of the droplet when determining the contact angle: we perform a circular fit through the top part of the spherical cap, and extrapolate towards the substrate (which is defined to be $\sigma/2$ above the top row of substrate atoms) to find $\theta$ and $R$. Fig.~\ref{fig:collapse}(a) shows these fits through some isodensity contours ($\rho^\ast=0.3,0.5,0.7$). 

This leads, however, to a new problem: which isodensity should one choose? As can readily be seen from Fig.~\ref{fig:collapse}(a) it turns out that the width of the interface cannot be neglected, and choosing different isodensity contours results in different values for the $\theta$ and $R$. To overcome this problem, we use the data from the cylindrical droplets to determine which isodensity contour to use. From a macroscopic perspective, the cylindrical caps are not affected by line tension ($\theta=\theta_Y$). It turns out that this property is obeyed by the Gibbs dividing surface at $\rho^\ast = 0.5$, where $\rho^\ast$ is a parametrized version of the local density given by:
\begin{equation}
\rho^\ast(\vec{r}) = \frac{\rho(\vec{r}) - \rho_V}{\rho_L-\rho_V}\;.
\end{equation}
Here, $\rho_L$ and $\rho_V$ are the bulk densities of the liquid and vapour phase, respectively.
We note that although line tension does not affect cylindrical droplets, other curvature effects (such as the Tolman-correction on $\gamma$ and the effect of the increased Laplace-pressure on $\gamma_{sl}$) \emph{do} play a role. The baseline established by this methodology is therefore not based on `pure' line tension, but rather an apparent line tension in which all these effects are combined. This is in line with previous experimental work, where this distinction could also not be made.

To improve statistics, we have also used the remaining isodensity contours. This can be done since the density profile across the interface is accurately fitted by (Fig.~\ref{fig:collapse}(b)):
\begin{equation}
\label{eq:dens}
\rho^\ast=\frac{1}{2}\left( 1+ \tanh \left( \frac{R_0-r}{w}\right) \right)\;,
\end{equation}
where $R_0$ is the point where the density is halfway between the liquid value and the vapour value ($\rho^\ast = 0.5$). $w$ is a fit parameter that defines the width of the liquid-vapour transition. Since the circular fits are concentric (they share a common center point, $\mathcal{C}$ in Fig.~ \ref{fig:collapse}(a)) we can easily transform any isodensity contour to the reference contour. For this, we calculate the radial distance from the contour towards the reference contour using \eqref{eq:dens}. The result of this transformation for the spherical droplet from Fig.~\ref{DropsMD} is shown in Fig.~ \ref{fig:collapse}(c), where we see that the contour shapes indeed collapse and can now all be used to determine the contact angle. The spread of these values for different $\rho^\ast$ are used to determine the error of the measurements. 

\subsection{Results: tension length}
\begin{figure}[htbp]
\begin{center}
\includegraphics[width=220pt]{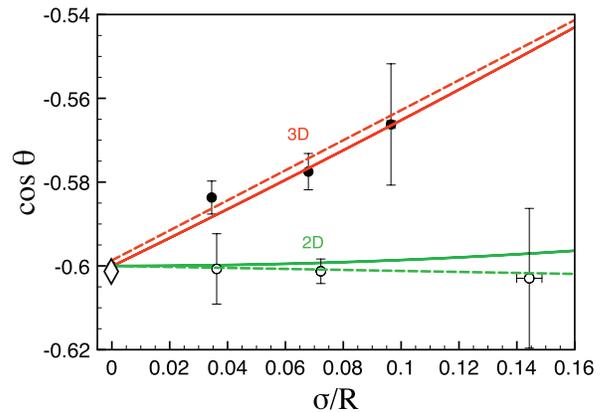}
\caption{$\cos \theta$ vs $\sigma R^{-1}$ for cylindrical (open symbols) and spherical (filled symbols) drops for $\cos\theta_{Y}=-0.60$ ($\approx 127^\circ$). The dashed lines are linear fits through the data points, and the solid lines are the solutions obtained using DFT described in section~\ref{sec:orig}. The top two (red) lines represent the 3D-data, whereas the bottom two (green) lines the 2D-data. The diamond at $1/R=0$ indicates Young's law, calculated independently by determining the surface tensions of the three interfaces: $\gamma$, $\gamma_{SL}$, $\gamma_{SV}$. The difference between the slopes of the 2D and 3D fits quantifies the tension length $\ell$. Note that for this particular equilibrium contact angle, MD and DFT agree quantitatively on the tension length.}
\label{ThetaMDsharp}
\end{center}
\end{figure}
\begin{figure}[htbp]
\begin{center}
\includegraphics{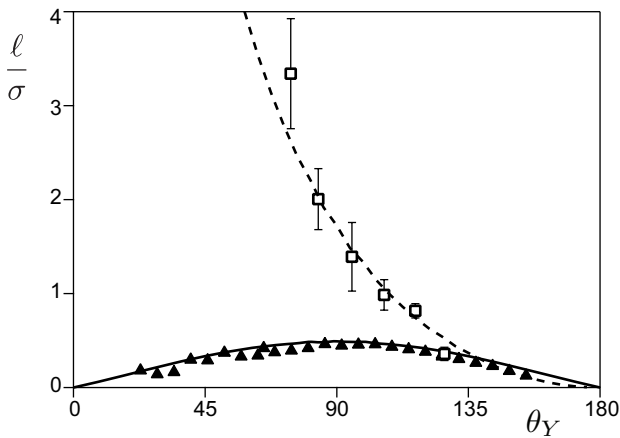}
\caption{Tension length $\ell$ vs $\theta_Y$ for spherical drops. Square symbols are the molecular dynamics results, triangle symbols are the results from the self-consistent Density Functional Theory model discussed in section \ref{subs:DFT}. These data points were acquired by measuring the contact angle for different drop sizes, meaning they represent an `apparent' line tension. The solid and dashed lines also result from DFT, assuming a wedge-shaped geometry near the contact line: eqs.~\eqref{eq:tauwedgeLS}, \eqref{eq:tauwedgeLLsmalltheta}, \eqref{eq:tauwedgeLL}. For the self-consistent DFT data the characteristic lengths are determined analytically: $\zeta_{LL}=\zeta_{LS}=\pi \sigma/4$. The resulting curve is the solid line. The characteristic lengths for the MD data are acquired by fitting: $\zeta_{LL}=3.5\,\sigma$, $\zeta_{LS}=0$, represented by the dashed line.}
\label{TensionLength}
\end{center}
\end{figure}
\begin{table}
\begin{tabular}{c|c|c|c}
$\epsilon_{LS}/\epsilon_{LL}$ & $\theta_Y$ & $\theta_\textrm{DFT}$ & $\ell$ ($\sigma$) \\ \hline 
\hline 0.33 & $129 \pm 1.0^\circ$ &  $109^\circ$& $0.36\pm 0.02$ \\ 
\hline 0.40 & $117 \pm 1.2^\circ$ &  $102^\circ$& $0.82\pm 0.03$ \\ 
\hline 0.47 & $106 \pm 1.3^\circ$ &  $94^\circ$& $0.99\pm 0.09$ \\ 
\hline 0.53 & $95 \pm 1.8^\circ$ &  $86^\circ$&  $1.39\pm 0.20$\\ 
\hline 0.60 & $84 \pm 1.5^\circ$ &  $78^\circ$&  $2.01\pm 0.18$\\ 
\hline 0.67 & $74 \pm 1.8^\circ$ &  $71^\circ$&  $3.34\pm 0.32$\\ 
\end{tabular} 
\caption{\label{tbl:MD}MD results as shown in Fig.~\ref{TensionLength}. The ratio $\epsilon_{LS}/\epsilon_{LL}$ was varied to obtain the tension length $\ell$ for different equilibrium contact angles $\theta_Y$. $\theta_\textrm{DFT}$ is the contact angle resulting from the interaction ratio according to the DFT-model described in Sect.~\ref{sec:orig}.}
\end{table}

Figure~\ref{ThetaMDsharp} shows the relation between the contact angle defined as described above and the drop radius for different sized droplets. Young's angle $\theta_Y$ was independently calculated from independent measurements of the surface tensions~\cite{Nijmeijer} of planar interfaces, under the same simulation conditions as the droplet simulations. This is shown as the diamond symbol at $1/R=0$. One can observe that the contact angle does not present any variation with drop size in the cylindrical cap case, which is consistent with the macroscopic picture. By contrast, the spherical drops exhibit a decreasing contact angle for small radii (large $R^{-1}$), which according to (\ref{eq:modyoung}) is consistent with a negative line tension $\tau$. A negative value of $\tau$ means that the contact line has the tendency to expand -- a larger contact line length leads to a \emph{decrease} in $\theta$ under the constant volume constraint. The solid line corresponds to the  density functional theory in the sharp kink approximation that will be discussed below. 

The difference between the slopes of the 2D- and 3D-fits in Fig.~\ref{ThetaMDsharp} (dashed lines) is equal to the tension length, $\ell\equiv -\tau/\gamma$, which is defined to be positive for negative values of $\tau$ (see \eqref{eq:ell}). For this equilibrium contact angle ($\theta_Y=127^\circ$) we find $\ell=0.36\sigma$. Now, by varying the interaction ratio $\epsilon_{LS}/\epsilon_{LL}$ we measure $\ell$ for varying $\theta_Y$. The result is shown in table~\ref{tbl:MD} and in Fig.~\ref{TensionLength} by the square symbols. Whatever $\theta_Y$, the tension length turns out to be positive (so $\tau$ is always negative) and very small --~of the order of the atomic size $\sigma$. The tension length recovered from the MD simulations is a decreasing function of $\theta_Y$, indicating that the effect is stronger when the wedge formed by the liquid in the vicinity of the contact line is sharp. The other curves in Fig.~\ref{TensionLength} result from DFT, and will be discussed in the following sections.

\section{Origin of line tension effect}
\label{sec:orig}
In this section we study line tension in the framework of Density Functional Theory using the sharp kink approximation. Once more, the strategy is to determine the equilibrium shapes of 2D and 3D drops and to compare their contact angles. Starting from the basic equations of DFT we first motivate the form of the free energy functional in Sec.~\ref{subs:DFT}. Some of the assumptions are directly tested using Molecular Dynamics simulations. We then derive the equilibrium condition for the capillary pressure (Sec.~\ref{subs:capp}) and describe the numerical scheme that was used to solve the equilibrium shapes of the drops (Sec.~\ref{subs:shaperel}). The numerical results are presented and interpreted in detail in Secs.~\ref{subs:results} and~\ref{subs:geom}.
\subsection{Density Functional Theory in the sharp kink approximation}
\label{subs:DFT}
The primary idea of Density Functional Theory (DFT) is to express the grand potential $\Omega=U-TS-\mu N=F-\mu N$ as a functional of the particle density $\rho$ and to perform a functional minimisation for a given $\mu$ and $T$. 
For an ideal gas, the free energy functional is known explicitly
\begin{equation}
F_{id}[\rho]=kT\int \rho\left[\ln(\rho \Lambda^3)-1\right] d\vec r\;,
\end{equation}
but this is not the case for general liquids. Let us denote $\phi(\vec r_1,\vec r_2)$ or $\phi(r)$ as the additive pair potential between particles at $\vec r_1$ and $\vec r_2$ with distance $r=|\vec r_2 - \vec r_1|$. From a grand canonical averaging, one can show (see e.g. refs. \onlinecite{Rowlinson/Widom,Hansen})
\begin{equation}
\frac{\delta \Omega}{\delta \phi(\vec r_1,\vec r_2)}= \frac{1}{2} \rho^{(2)}(\vec r_1,\vec r_2)= \frac{1}{2} \rho(\vec r_1)\rho(\vec r_2) g(\vec r_1,\vec r_2)~,
\label{eqG}
\end{equation}
where $\rho^{(2)}$ is the two-body density distribution function and $g$ the pair correlation function. This relation can be used to construct the free energy for non-ideal systems. Introducing a coupling parameter $\lambda$ in front of the interaction, the free energy can be constructed by integration as:
\begin{eqnarray}
F[\rho]=F_{id}[\rho] + \nonumber \\
\frac{1}{2} \int_0^1 d\lambda \int d\vec r_1\int d\vec r_2 \rho(\vec r_1)\rho(\vec r_2) g_\lambda(\vec r_1,\vec r_2)\phi(|\vec r_2-\vec r_1|)~.\nonumber \\
\label{eqF9}
\end{eqnarray}
Here $g_\lambda$ is the pair correlation function in a system of same geometry and same volume, for which the interaction is $\lambda\phi(r)$.

Although exact, this expression cannot be used as it is, as $g_\lambda$ is not known. For a practical approximation of the energy functional, one can separate the thermodynamic non-ideality in contributions due to attractive and repulsive components of the intermolecular potential. As the repulsive forces have a very short range, their effect is mainly local. Using the Local Density Approximation, the repulsive contribution can be estimated from the Helmoltz energy density $f_r(\rho)$ in a uniform system of density $\rho$ at temperature $T$, composed of purely repulsive molecules. The attractive Van der Waals interactions, $\phi_{att}$, can then be treated as a perturbation, assuming that the pair correlation function remains mostly that of the purely repulsive reference system, $g_r(\vec r_1,\vec r_2)$. The free energy then reads:
\begin{eqnarray}
F[\rho]=\int f_r(\rho) d\vec r+ \nonumber \\
\frac{1}{2} \int d\vec r_1\int d\vec r_2 \rho(\vec r_1)\rho(\vec r_2) g_r(\vec r_1,\vec r_2)
\phi_{att}(|\vec r_2-\vec r_1|)~. \nonumber \\
\label{eqF10}
\end{eqnarray}
\begin{figure}[tbp]
\begin{center}
\includegraphics{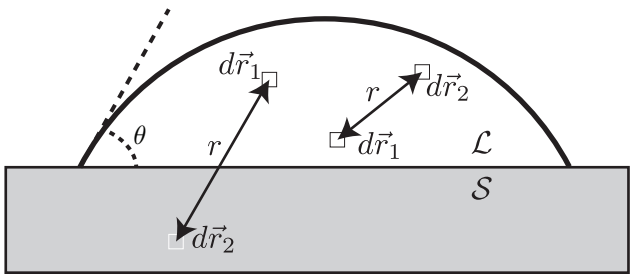}
\caption{Schematic representation of the integration variables and domains from \eqref{eq:F}. Both  the liquid-solid (left) and the liquid-liquid (right) interactions are integrated over their respective volumes (the liquid cap $\mathcal{L}$ and the solid substrate $\mathcal{S}$) to obtain the total free energy $F$.}
\label{fig:f_integral}
\end{center}
\end{figure}

To end up with a numerically tractable scheme, we make a final approximation that the density profile across the interface is mostly independent of the geometry. Defining the position of the interface e.g. by the iso-density $\rho^*=1/2$, the integrals in~(\ref{eqF10}) can be approximated by assuming that the density is uniform in both phases~\cite{GettaDietrich}.This so-called `sharp kink approximation' neglects the thickness of the diffuse interface. Thermal effects are implicitly taken into account, since $f_r$, $g_r$ and the liquid density depend on temperature. In this approximation, the free energy becomes an explicit functional of the shape of liquid, solid, and vapor domains. Since the vapor density is neglible with respect to that of the solid and liquid, we find
\begin{eqnarray}\label{eq:F}
F &=& f_r(\rho_L) \int_{\cal L}  d\vec r + f_r(\rho_S) \int_{\cal S}  d\vec r \nonumber \\
&+&\frac{1}{2} \rho_L^2 \int_{\cal L} d\vec r_1\int_{\cal L} d\vec r_2 g_r(|\vec r_2-\vec r_1|)\phi_{LL}(|\vec r_2-\vec r_1|)  \nonumber \\
&+&\rho_L \rho_S  \int_{\cal L} d\vec r_1\int_{\cal S} d\vec r_2 g_r(|\vec r_2-\vec r_1|)\phi_{LS}(|\vec r_2-\vec r_1|)~,  \nonumber \\
&&
\label{eqF11}
\end{eqnarray}
where ${\cal L}$ and ${\cal S}$ are the liquid and solid domains respectively, see also Fig.~\ref{fig:f_integral}. $\phi_{LL}$ and $\phi_{LS}$ denote the attractive parts of the respective interactions. The drop shapes are found by minimizing this free energy with respect to the shape of the liquid domain ${\cal L}$. 

\begin{figure}[htbp]
\begin{center}
\includegraphics{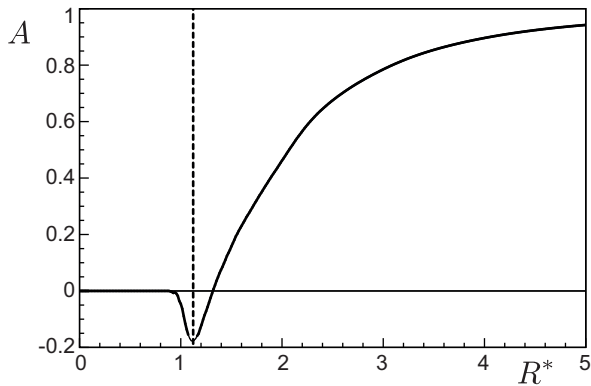}
\caption{Stress anisotropy $A$ for bond lengths smaller than $R^\ast$, see (\ref{eq:anisvr}). The measurement was done in a slab of height $\sigma/3$ within the liquid-vapour interface in a molecular dynamics simulation. The dashed line indicates the minimum of the Lennard-Jones potential at $R^\ast = 2^{1/6}\sigma$ and marks the separation between the attractive and repulsive bonds. The majority of the anisotropy comes from attraction.}
\label{fig:stressanis}
\end{center}
\end{figure}

Before proceeding, it is instructive to discuss the approximations underlying (\ref{eqF11}) in the light of the Molecular Dynamics simulations of the Lennard-Jones droplets. First, the assumption that the pair-correlation function is homogeneous in space ignores the layering near the solid wall (cf. Fig.~\ref{DropsMD}). This can induce significant corrections to the estimated free energy. Second, the Local Density Approximation of the short-range repulsive forces gives rise to isotropic repulsive interactions, while the attractive interactions will become anisotropic in the vicinity of an interface. If this is indeed the case, the surface tension (and line tension) effects mostly result from the attractive component of the interaction. We test the validity of this hypothesis in the Molecular Dynamics simulations by measuring the anisotropy of the stress tensor in the vicinity of the liquid-vapour interface. We define a cumulative stress-tensor $\sigma^{\alpha \alpha}(R^\ast)$ that incorporates only the interactions with a bond length smaller than $R^\ast$:

\begin{equation}
\label{eq:rstresst}
\bar{\bar{\sigma}}^{\alpha \alpha}({R^\ast}) =  \sum_i {m_i v_i^\alpha v_i^\alpha} - \sum_{j \ne i} \sum_{|r_{ij}|<R^\ast}{f^{\alpha}_{ij} r^\alpha_{ij}}\;.
\end{equation}
The true stress in the system is recovered when $R^\ast=\infty$, for which all interactions are taken into account. Here, $m_i$, $v_i$ are the mass and velocity of particle $i$, respectively, and $f_{ij}$ and $r_{ij}$ are the force and displacement vector between particles $i$ and $j$. With this, we quantify the anisotropy from the difference between the stress components tangential ($T$) and normal ($N$) to the interface, as

\begin{equation}
\label{eq:anisvr}
A(R^\ast)=\frac{ \bar{\bar{\sigma}}^{TT}(R^\ast) - \bar{\bar{\sigma}}^{NN}(R^\ast) } { \bar{\bar{\sigma}}^{TT}(\infty) + \bar{\bar{\sigma}}^{NN}(\infty) }\;.
\end{equation}
Figure~\ref{fig:stressanis} shows the anisotropy $A$ as a function of $R^\ast$. The dashed line indicates the transition from the repulsive ($r<2^{1/6}\sigma$) to the attractive domain ($r>2^{1/6}\sigma$). The figure clearly shows that the majority of the anisotropy in the liquid-vapour interface is due to the attractive interaction, while the repulsive interaction accounts for about 20\% of the anisotropy. This indeed justifies a local density approximation for the repulsion, although one can expect quantitative differences with Molecular Dynamics.

\subsection{Capillary pressure}
\label{subs:capp}
The equilibrium shape of liquid drops can be obtained by minimizing the free energy $F$ at constant volume $V$. This can be done by variation of (\ref{eqF11}) with respect to the drop shape $\mathcal{L}$ under the constraint of constant volume. The resulting equilibrium condition is a constant potential energy density $\Pi$ along the free surface~\cite{MerchantKeller, Snoeijer,GettaDietrich}. This potential can be interpreted as the capillary pressure and can be decomposed into a liquid-liquid and a solid-liquid contribution, as $\Pi=\Pi_{LL} + \Pi_{LS}$. The former can be written as
\begin{equation}
\Pi_{LL}\left(\vec r \right)=-\Pi_{LL}^\circ +  \rho_L\int_{\cal L} d\vec r' g_r(|\vec r'-\vec r|)\phi_{LL}(|\vec r'-\vec r|)\;,
\label{eqpill}
\end{equation}
where we subtracted $\Pi_{LL}^\circ$, the interaction due to a semi-infinite volume of liquid. The solid-liquid contribution follows from the interaction due to the semi-infinite volume of solid
\begin{equation}
\Pi_{LS}\left(\vec r \right)=\rho_S\int_{\cal S} d\vec r' g_r(|\vec r'-\vec r|)\phi_{LS}(|\vec r'-\vec r|)\;.
\end{equation}
The equilibrium condition is thus that 

\begin{equation}
\Pi \left(\vec r^* \right) =\Pi_{LL} \left(\vec r^* \right) + \Pi_{LS} \left(\vec r^* \right)  = {\rm constant},
\end{equation}
where $\vec r^*$ denotes an arbitrary position at the liquid-vapor interface.

Note that the capillary pressure $\Pi$ depends on the shape of the liquid, through the domain of integrations, and thus reflects the effect of the interface geometry on the free energy. It implicitly contains the Laplace pressure, which is the capillary pressure associated to a macroscopic curvature of the interface, and the disjoining pressure, which is the capillary pressure  in the case of a microscopic film.

\subsection{Shape relaxation}
\label{subs:shaperel}
We compute droplet shapes for a pair interaction consistent with the long-range van der Waals interaction used in the Molecular Dynamics simulations (see Fig.~\ref{fig:pots}):

\begin{equation}
\label{eq:regpot}
\rho_{i}\rho_{j}g_r(r)\phi_{ij}(r)=\frac{-c_{ij}}{\left(\sigma^2+r^2\right)^3}\;,
\end{equation}
where $c_{ij}$ represents the strength of the interaction between molecule $i$ and $j$. Comparing to the van der Waals interaction of (\ref{eq:LJ}), one finds $c_{ij}=4\rho_i \rho_j \sigma^6 \epsilon_{ij}$. For mathematical convenience we have chosen a simple regularization around $r =\sigma$, which represents the effective size of the short-range repulsion. Let us note that the potential of equation (17) does not lead to the formation of a precursor film. Namely, the corresponding energy per unit surface for a flat film is a monotonic function of the thickness $h$, with a prefactor depending on the spreading parameter. For partial wetting, the system tends to zero thickness rather than to a  precursor film of finite thickness. We have tested that other similar choices like $g(r)=0$ for $r<\sigma$ and $g(r)=1$ for $r>\sigma$ leads to quantitatively similar results~(see ref. \onlinecite{Snoeijer}). The surface tensions corresponding to (\ref{eq:regpot}) can be computed as 
\begin{equation}
\gamma = \pi c_{ll}/8\sigma^2\;,
\label{eq:gamma_mic}
\end{equation}
 and
 \begin{equation}
 \label{eq:gammas_mic}
\gamma+\gamma_{sv}-\gamma_{sl}  = \pi c_{ls}/4\sigma^2\;.
\end{equation} 
By choosing identical functional forms for both interactions, one simply has~\cite{Rowlinson/Widom}
\begin{equation}\label{eq:thetaDFT}
\cos \theta_Y=\cos \theta_{ DFT} \equiv 2\frac{c_{ls}}{c_{ll}} - 1.
\end{equation}
\begin{figure}[t!]
\begin{center}
\includegraphics{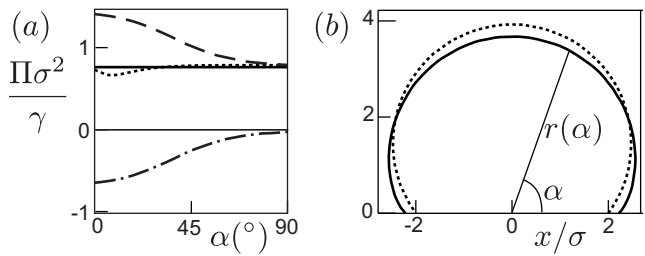}
\caption{\label{fig:dftpot}Typical result of the DFT model in the sharp kink approximation. (a) Surface potentials of an axisymmetric drop with $\theta_{Y}=127^\circ$, a measured angle $\theta=116^\circ$ and radius $R\approx 2\sigma$. $\alpha=0$ at the contact line. The dashed line (top) represents $\tilde\Pi_{ll}$, the bottom line represents $\Pi_{sl}$, and the lines in the middle indicate the sum of the two. Here, the dashed and solid lines indicate the potential energy density of the droplet in its initial shape, and its equilibrium (final) shape, respectively. (b) Initial and final drop profiles shown by the dashed and solid lines, respectively. The initial profile is a spherical cap shaped drop with $\theta=\theta_Y=127^\circ$.}
\end{center}
\end{figure}

Similar to the Molecular Dynamics simulations, we compute the equilibrium shapes of nanodrops in both the 2D configuration (cylindrical caps) and 3D configuration (spherical caps). The drop shapes are parameterized by $r(\alpha)$ as shown in Fig.~\ref{fig:dftpot} --~ polar coordinates are used to allow contact angles larger than $\pi/2$. We numerically determine the equilibrium shape of the drop by an iterative algorithm that tends to a constant $\Pi(\alpha)$ along the interface. The initial shape is taken as a spherical cap with $\theta_Y$ according to (\ref{eq:thetaDFT}). This is shown in Fig.~\ref{fig:dftpot} by dashed lines. The corresponding potential $\Pi(\alpha)$ is uniform except within a few molecular scales from the contact line, where the influence of the solid plays a role. We iteratively construct drop shapes $r^t(\alpha)$ according to $r^{t+1}(\alpha)=r^t(\alpha)+\lambda^t (\pi^t(\alpha)-\langle\pi^t\rangle_{\alpha})$, while keeping the volume constant. Here, $\pi^t(\alpha )$ is the capillary pressure at angle $\alpha$ during iteration $t$. $\left< \pi^t \right>_{\alpha}$ is the space-averaged potential at the interface during iteration $t$. The parameter $\lambda^t$ is selected such that the variance of the potential is minimised at each step. After a few hundred steps, the shape converges and yields $\Pi(\alpha)$ that indeed is constant within numerical precision. Note that for the potential studied here, no precursor film is formed.
\begin{figure}[t!]
\begin{center}
\includegraphics{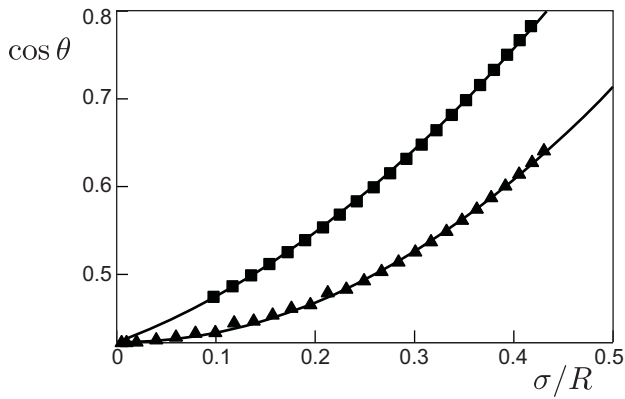}
\caption{\label{fig:DFTplot}Cosine of the equilibrium contact angle against $1/R$, for 2D drops (triangles) and 3D drops (squares). The corresponding Young's angle is $\theta_{Y}=65^\circ$. The slope of the curve near $1/R=0$ can be attributed to line tension for the 3D drops, while it is zero for 2D drops. Note that both curves exhibit a significant $1/R^2$ contribution for smaller drop sizes. This contribution was not recovered from the molecular dynamics simulations, since the radius of the droplets was not small enough: $R>7\sigma$. Smaller droplets would not allow for spherical cap fitting because the droplet size becomes similar to the particle size.}
\end{center}
\end{figure}

The shape $r(\alpha)$ and the capillary pressure $\Pi(\alpha)$ of a small drop are plotted as solid lines in Fig.~\ref{fig:dftpot}. Away from the contact line the drop is a spherical cap, but a significant deviation can be observed near the contact line. The drop has spread with respect to the initial shape, resulting into a lower contact angle than $\theta_Y$. Once more, this is consistent with a negative value of the line tension $\tau$. Far from the contact line, the capillary pressure is dominated by the $\Pi_{ll}$ term. The corresponding value is simply the expected Laplace pressure $2\gamma /\tilde{R}$, where $\tilde{R}$ is the radius of curvature the drop.

\subsection{Results}
\label{subs:results}
\begin{figure}[t!]
\begin{center}
\includegraphics{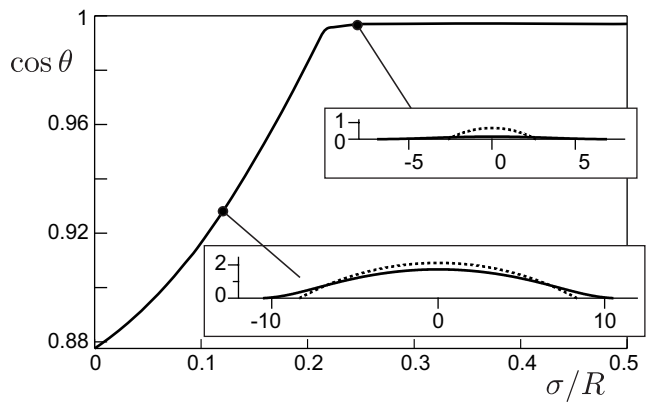}
\caption{\label{fig:spreadingeffect}Cosine of the equilibrium contact angle against $1/R$, for a 3D drop $\theta_{Y}=28^\circ$ as shown in the insets for two different sizes $R$. At sufficiently small radius we observe a saturation of $\cos(\theta)=1$, approaching a perfectly wetting drop (top inset).}
\end{center}
\end{figure}

In Fig.~\ref{fig:DFTplot} we compare the contact angles of 3D drops (squares) and 2D drops (triangles), as a function of inverse drop radius $1/R$. In both cases the interactions were identical, corresponding to $\theta_Y=65\degree$. For large drops there is indeed a difference that can be attributed to line tension: the slope at $1/R \rightarrow 0$ is finite for 3D drops while it vanishes in the 2D case. Interestingly, however, there remains a $1/R^2$ contribution for both types of drops. The two data sets are accurately fitted by parabola, with equal prefactors for the quadratic term. This suggests that the effect of line tension in (\ref{eq:modyoung}) can be seen as the leading order contribution of an expansion in $\sigma/R$, and is only valid for relatively large drops. In particular, (\ref{eq:modyoung}) must break down when $\cos \theta \approx 1$. This is illustrated by Fig.~\ref{fig:spreadingeffect} showing a saturation of the contact angle to $\theta\approx 0$ for very small drops. This effect is of course most pronounced for drops that already have a small Young's angle $\theta_Y$. For such small drops, the range over which one observes a $1/R$ behaviour is very small and the main size effect is to induce a wetting transition.

We are now in the position to make a comparison of the DFT model with the molecular dynamics simulations presented in Sec.~\ref{sec:MD}. The solid line in Fig.~\ref{ThetaMDsharp}  represents the contact angles for 3D drops of varying sizes as obtained from the numerical DFT calculation. We took the same Young's angle as obtained in the molecular dynamics simulations, i.e. $\theta_{Y}=127\degree$. The trends of DFT and molecular dynamics are very similar, clearly showing a decrease in contact angle for decreasing drop radius. For both cases line tension is thus negative and has a similar magnitude in units of $\sigma$. 

Finally, we determined the tension length $\ell$ from the slope near $1/R\rightarrow 0$, for a broad range of $\theta_Y$. The results are reported as triangles in Fig.~\ref{TensionLength}. The value of $\ell$ vanishes both for $0$ and $180\degree$ and presents a maximum around $90\degree$. Beside the sign and the order of magnitude, the behaviour is thus qualitatively different from the molecular dynamics results. This difference is most pronounced at small $\theta_Y$ (Fig.~\ref{TensionLength}).

\subsection{Geometric interpretation of line tension}
\label{subs:geom}

\begin{figure}[t!]
\begin{center}
\includegraphics{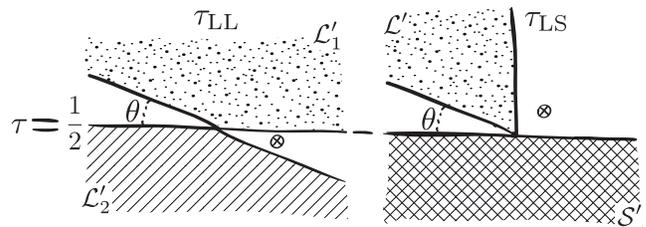}
\caption{Integration domains of the free energy~\eqref{eq:F} that contribute to line tension due to liquid-liquid interactions $\tau_{LL}$ (left) and solid-liquid interactions $\tau_{LS}$ (right). Assuming the contact line region to be perfectly wedge-shaped the total free energy $F$ minus the volumic and surfacic contributions results in a residual energy, which can be attributed to line tension. We show in the Appendix how the line tension contribution can be isolated from the liquid-liquid and solid-liquid interactions. Then, by calculating the free energy associated with these integration domains (Eq.~\eqref{eq:ltcomp}) one directly finds the line tension due to liquid-liquid interactions and solid-liquid interactions: eqs.~\eqref{eq:tauwedgeLS},~\eqref{eq:tauwedgeLLsmalltheta}, and~\eqref{eq:tauwedgeLL}. }
\label{fig:linetension}
\end{center}
\end{figure}

Within our DFT model, the dependence of $\ell$ on contact angle $\theta_Y$ can be accurately described from a geometric argument (solid line in Fig.~\ref{TensionLength}). We separate the free energy \eqref{eq:F} in volumic, surfacic and linear contributions, as $F=PV+ \sum_i \gamma_i S_i+\tau L$. By assuming the liquid domain to be wedge-shaped, it is indeed possible to explicitly separate the domains of integration in \eqref{eq:F} in bulk, surface, and line  contributions:
\begin{eqnarray}
\label{eq:ltcomp}
\nonumber
F&=&PV+\sum_i \gamma_i S_i \\\nonumber
&+&\frac{1}{2} \rho_L^2 \int_{\mathcal{L}_1^\prime}d\vec r_1 \int_{\mathcal{L}_2^\prime} d \vec r_2 g_r(|\vec r_2 - \vec r_1|)\phi_{LL}(|\vec r_2 - \vec r_1|)\\
&+&\rho_L\rho_S \int_{\mathcal{L}^\prime}d\vec r_1 \int_{\mathcal{S}^\prime} d \vec r_2 g_r(|\vec r_2 - \vec r_1|)\phi_{LS}(|\vec r_2 - \vec r_1|)  \;.\nonumber \\
&&
\end{eqnarray}
Here the integration domains $\mathcal{L}_1^\prime$, $\mathcal{L}_2^\prime$, $\mathcal{L}^\prime$, and $\mathcal{S}^\prime$ are those represented in Fig.~\ref{fig:linetension} and the Appendix. Note that such a decomposition is uniquely defined in the sharp-kink approximation, while this is no longer the case for inhomogeneous density profiles.

 From \eqref{eq:ltcomp} one sees directly that line tension has two contributions, due to liquid-liquid interactions $\tau_{LL}$ , and due to liquid-solid interactions $\tau_{LS}$. These can be computed as follows. The integration domains $\mathcal{L}_1^\prime$, $\mathcal{L}_2^\prime$, $\mathcal{L}^\prime$, and $\mathcal{S}^\prime$ are bordered by straight lines passing through the contact line so that they do not present a characteristic scale. Therefore, both $\tau_{LL}$ and $\tau_{LS}$ can be written as products of a characteristic length (that does not depend on $\theta$), and a function of $\theta$ (that does not depend on the potentials $\phi_{LL}$ and $\phi_{LS}$). It turns out that the lengths can be expressed in terms of  the liquid-liquid and solid-liquid disjoining pressures $\Pi^{\rm disj}_{LL}(h)$ and $\Pi^{\rm disj}_{ LS}(h)$. The disjoining pressure is the energy per unit liquid volume at a distance $h$ from a flat semi-infinite zone of liquid or solid (see the integration domain $\mathcal{S}^\prime$ in Fig.~\ref{fig:linetension}). The surface tensions, already computed in (\ref{eq:gamma_mic},\ref{eq:gammas_mic}), can be expressed as the integrals of these quantities:
\begin{eqnarray}
\int \Pi^{\rm disj}_{ LL}(h)\,dh=2\gamma\\
\int \Pi^{\rm disj}_{ LS}(h)\,dh=\gamma+\gamma_{sv}-\gamma_{sl} =\gamma\;(1+\cos\theta_Y)
\end{eqnarray}
The characteristic lengths $\zeta_{LL}$ and $\zeta_{LS}$ that appear in the calculation of the line tension turn out to be the first moment of the disjoining pressure:
\begin{equation}
\zeta_{LL}=\frac{\int z\,\Pi^{\rm disj}_{LL}(z)\,dz}{\int \Pi^{\rm disj}_{LL}(z)\,dz}\quad{\rm and}\quad \zeta_{  LS}=\frac{\int z\,\Pi^{\rm disj}_{LS}(z)\,dz}{\int \Pi^{\rm disj}_{ LS}(z)\,dz}
\end{equation}
Following the interpretation of surface tension as a force per unit length, $\zeta_{LL}$ and $\zeta_{  LS}$  are the ``moment arms'' of these forces. Within our DFT model, the liquid-liquid and solid-liquid potentials have the same shape so that these two lengths are equal, $\zeta_{ LS}=\zeta_{LL}=\pi \sigma/4$. 

The line tension $\tau_{LS}$ follows as:
\begin{equation}
\label{eq:tauwedgeLS}
\tau_{LS} = \zeta_{ LS}\gamma\;\frac{(1+\cos\theta_Y)}{\tan\theta}.
\end{equation}
This contribution is positive for $0<\theta<\pi/2$ and changes sign at $\theta=\pi/2$. The prefactor $(1+\cos \theta_Y)$ is not of geometric origin, but stems from the strength of the liquid-solid interaction $c_{LS}$. A similar result for $\tau_{LS}$ was previously obtained in ref. \onlinecite{Marmur}, but this work omitted the contribution due to liquid-liquid interactions, $\tau_{LL}$, which is crucial to describe our numerical DFT results. The contribution due to liquid-liquid interactions is negative for all angles.  In the limit of small angles, $\tau_{LL}$ diverges as
\begin{equation}
\label{eq:tauwedgeLLsmalltheta}
\tau_{LL} \equiv -\zeta_{LL}\;\gamma\;\frac{2}{\tan\theta}.
\end{equation}
As the angle $\theta$ tends to $\pi$, $\tau_{LL}$  vanishes as
\begin{equation}
\label{eq:tauwedgeLL}
\tau_{LL} = - \frac{2}{3\pi}\,\zeta_{LL}\;\gamma  \left(\pi-\theta\right)^2
\end{equation}
In between, we have determined the ratio $\tau_{LL}/\zeta_{LL}$ by numerical integration. 

Adding the two contributions $\tau_{LL}$ and $\tau_{LS}$, we obtain the solid line in Fig.~\ref{TensionLength}, which indeed closely follows the full numerical simulations obtained from the spherical cap measurements. Note that both $\tau_{LL}$ and $\tau_{LS}$ scale as $1/\theta$ for small angles, but the diverging contributions balance exactly. This is a consequence of having identical values for the moment arms, i.e. $\zeta_{LL}=\zeta_{LS}$, resulting in a vanishing line tension for small $\theta$. Of course, this will not be the case in general, where we expect one of the contributions to dominate.

\section{Discussion}

We theoretically investigated the effect of line tension by studying the contact angles of Lennard-Jones droplets of varying sizes. The equilibrium shapes of nanodrops were determined using two methods: Molecular Dynamics (MD) and Density Functional Theory (DFT). For 3D drops we found a size-dependent contact angle consistent with \eqref{eq:modyoung}, while the contact angle was nearly constant for 2D drops. DFT in the employed approximation does not fully reproduce the MD simulations, but it does capture the main physics. In particular, DFT gives the correct (negative) sign and order of magnitude of $\tau$, and also captures the dependence on wettability for large contact angles. Obiously, the exact numerical values resulting from the DFT calculation depend on our specific choice of the potential~\eqref{eq:regpot}. Note however, that both the recovered trend and the orders of magnitude for $\ell$ are a \emph{general} result, independent of the specific choice of the potential. The only exception is the limit of the wetting transition, $\theta \rightarrow 0$, which is known to depend on details of the interaction~\cite{Dobbs/Indekeu,Indekeu1994,Indekeu}.

In addition, we identified a simple geometric interpretation of line tension. Molecules inside a liquid wedge interact with a larger number of surrounding molecules than estimated from surface tension, which is based on an infinite half space of liquid. Hence the negative sign of line tension. The wedge shape of the liquid is indeed a good approximation of the liquid geometry for large contact angles and yields a very accurate prediction for $\tau$ in the DFT case. This is remarkable, since these DFT-measurements did not discriminate between line tension and other curvature effects, suggesting that line tension is the dominant mechanism for the size dependence of the contact angle. Once more, the behavior for small contact angles is sensitive to details of the interaction: it depends on the ``moment arm" of the surface tensions, characterized by the length scales $\zeta_{LL}$ and $\zeta_{LS}$. We speculate that the layering effect near the substrate in MD substantially reduces the moment arm $\zeta_{LS}$ for the liquid-solid interaction. This would explain the discrepancy with DFT. Indeed, the MD data can be described by the wedge-approximation of $\tau_{LL}$ by fitting the moment arms to $\zeta_{LL}=3.5 \sigma$ and $\zeta_{LS}=0$. It would be interesting to further investigate this matter.

Although we were able to observe the variation of contact angle with drop size, the effect is only noticeable for very small, nano-scale drops. Taking $\sigma=0.34$~nm and $\gamma=0.017$~J/m$^2$, our results correspond to line tension in the range $\tau=10^{-12}-10^{-11}$~J/m (depending on the wettability). This is consistent with theoretical predictions as well as with recent experiments~\cite{Pompe}. Note, however, that much larger experimental values for $\tau$ have also been reported~\cite{Gaydos,Li1990,Vera-Graziano}. Resolving this issue is particularly important for surface nanobubbles~\cite{BorkentPRL,Nakabayashi,Borkent,Hampton}, typically $100$nm wide, whose stability was suggested to rely on an effective line tension~\cite{Brenner}.

\acknowledgements

We wish to thank Lyderic Bocquet for his private lecture on DFT. This work was sponsored by the Stichting Nationale Computerfaciliteiten (National Computing Facilities Foundation, NCF) for the use of supercomputer facilities, with financial support from the Nederlandse Organisatie voor Wetenschappelijk Onderzoek (Netherlands Organisation for Scientific Research, NWO). This work is part of the research programme of the Foundation for Fundamental Research on Matter (FOM), which is part of the Netherlands Organisation for Scientific Research (NWO).

\appendix
\begin{figure*}[t!]
\begin{center}
\includegraphics{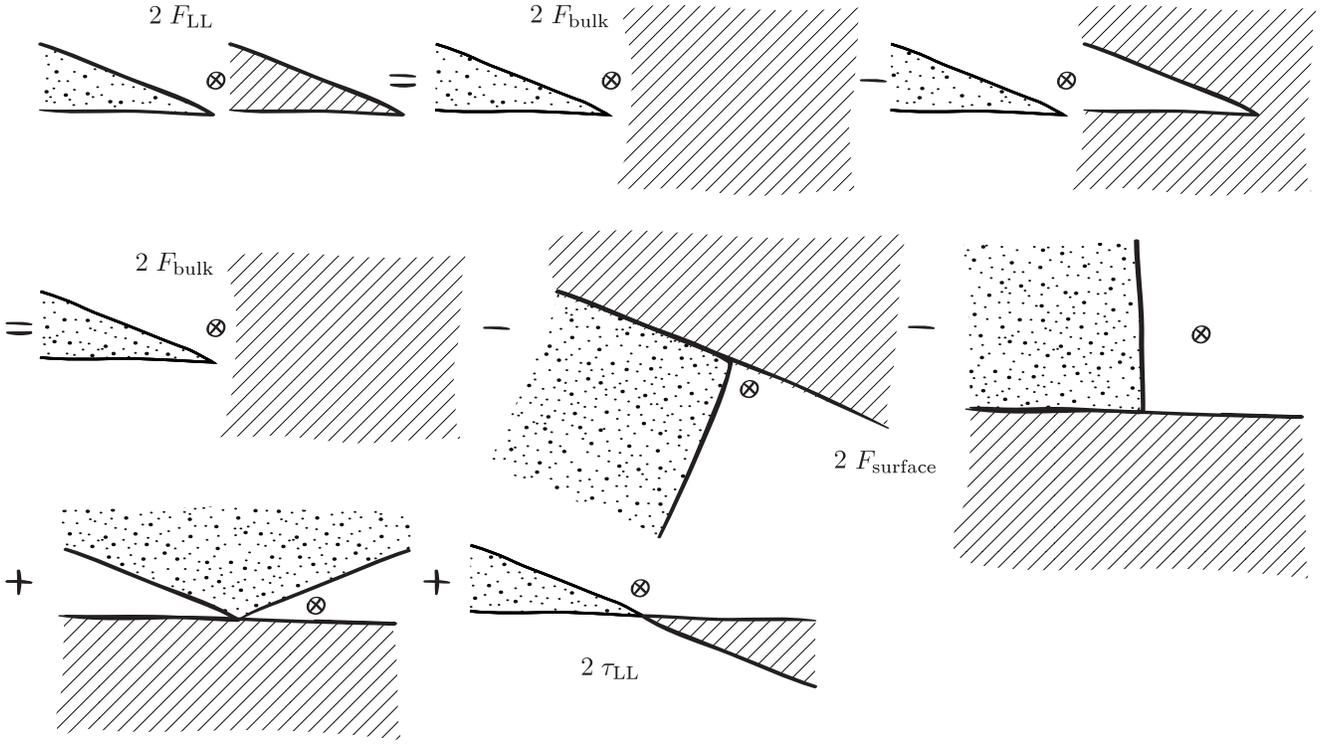}
\caption{Integration domain for the liquid-liquid interaction energy decomposed in the bulk, surface, and line components. The dotted and striped regions represent the domains of integration for the variables $d \vec r_1$ and $d \vec r_2$, respectively, in the liquid-liquid term of \eqref{eq:ltcomp}. The remainder after subtracting the bulk energy ($F_\textrm{bulk}$) and surface energy ($F_\textrm{surface}$) is the free energy associated with liquid-liquid line tension ($\tau_{LL}$). Note that the two line-tension contributions shown here can be combined into the integration domains shown in Fig.~\ref{fig:linetension}.}
\label{fig:geomLL}
\end{center}
\end{figure*}

\section{Geometric interpretion of line tension}
Figure~\ref{fig:geomLL} shows by illustration how the free energy associated with the liquid-liquid interactions of a wedge-shaped liquid interface sitting on a solid can be decomposed in its bulk, surface and contact line contributions. $F_{LL}$ shows the total free energy of the liquid-liquid interactions, which is the interaction of the liquid in the wedge with itself. For clarity, we separated the two domains in the first row spatially, but in reality they of course overlap since they are the same volume of liquid. First, we decompose the integral domain in the bulk energy contribution (wedge shape $\otimes$ infinite volume): $F_{\textrm{bulk}}$. The surplus that has to be subtracted is shown in the top right of Fig.~\ref{fig:geomLL}, because one has to compensate for the areas where no liquid is present. From this surplus we extract the surface contributions. Note that the liquid wedge has two surfaces: the liquid-vapour interface and the liquid-solid interface, which are both represented by integration of the wedge (dotted area) with an infinite half-space, resulting in the total surface energy term. The third row shows what remains and is by definition (Eq.~\eqref{eq:dF}) the line tension. These integration domains can be simplified and merged into the one shown in Fig.~\ref{fig:linetension} (left).

\begin{figure*}[h]
\begin{center}
\includegraphics{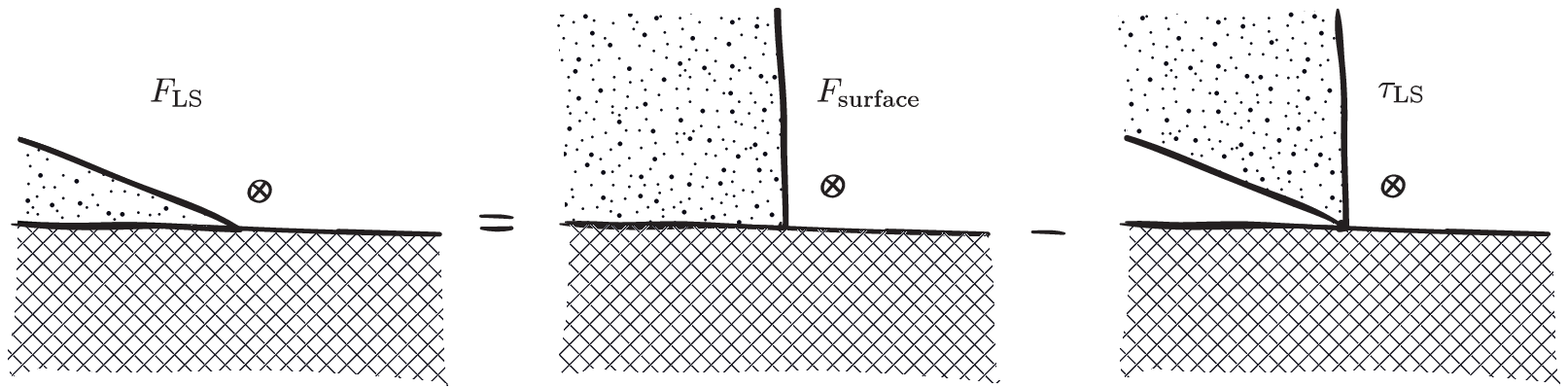}
\caption{Integration domain for the liquid-solid interaction energy (left). This integration domain can be decomposed in the corresponding surface energy contribution ($F_\textrm{surface}$) and the free energy associated with liquid-solid line tension ($\tau_{LS}$). The dotted region represents the integration variable $d\vec r_1$ in the liquid-solid term of \eqref{eq:ltcomp} and the double striped region the integration variable $d \vec r_2$ in the same equation.}
\label{fig:geomLS}
\end{center}
\end{figure*}

To compute $\tau_{LS}$ we follow a similar route. Fig.~\ref{fig:geomLS} (left) shows the integration domain for $F_{LS}$ for a liquid wedge (dotted) in contact with a solid (striped). The right two panels directly give the decomposition into the surfacic component (solid half-space $\otimes$ liquid half-space, over the solid-liquid interface), and the remainder which is the line tension component ($\tau_{LS}$). There is no bulk energy term since we are dealing with two separate (and spatially separated) phases.

\begin{thebibliography}{10}
\newcommand{\enquote}[1]{``#1''}
\expandafter\ifx\csname url\endcsname\relax
  \def\url#1{\texttt{#1}}\fi
\expandafter\ifx\csname urlprefix\endcsname\relax\def\urlprefix{URL }\fi
\providecommand{\bibinfo}[2]{#2}
\providecommand{\noopsort}[1]{}
\providecommand{\switchargs}[2]{#2#1}

\bibitem{Gibbs}
\bibinfo{author}{J.~W. Gibbs}, \emph{\bibinfo{title}{The Collected Works of J.
  Willard Gibbs}} (\bibinfo{publisher}{Yale University Press, London})
  (\bibinfo{year}{1957}).

\bibitem{Guzzardi}
\bibinfo{author}{L.~Guzzardi}, \bibinfo{author}{R.~Rosso}, and
  \bibinfo{author}{E.~G. Virga}, \enquote{\bibinfo{title}{Residual stability of
  sessile droplets with negative line tension}}, \bibinfo{journal}{Phys. Rev.
  E} \textbf{\bibinfo{volume}{73}}, \bibinfo{pages}{021602}
  (\bibinfo{year}{2006}).

\bibitem{SchimmeleEPJE}
\bibinfo{author}{L.~Schimmele} and \bibinfo{author}{S.~Dietrich},
  \enquote{\bibinfo{title}{Line tension and the shape of nanodroplets}},
  \bibinfo{journal}{Eur. Phys. J. E} \textbf{\bibinfo{volume}{30}},
  \bibinfo{pages}{427--430} (\bibinfo{year}{2009}).

\bibitem{SchimmeleJCP}
\bibinfo{author}{L.~Schimmele}, \bibinfo{author}{M.~Napiorkowski}, and
  \bibinfo{author}{S.~Dietrich}, \enquote{\bibinfo{title}{{Conceptual aspects
  of line tensions}}}, \bibinfo{journal}{{J. Chem. Phys.}}
  \textbf{\bibinfo{volume}{{127}}}, \bibinfo{pages}{{164715}}
  (\bibinfo{year}{{2007}}).

\bibitem{Pethica}
\bibinfo{author}{B.~Pethica}, \enquote{\bibinfo{title}{{Contact-Angle
  Equilibrium}}}, \bibinfo{journal}{{J. Colloid Interface Sci.}}
  \textbf{\bibinfo{volume}{{62}}}, \bibinfo{pages}{{567--569}}
  (\bibinfo{year}{{1977}}).

\bibitem{GettaDietrich}
\bibinfo{author}{T.~Getta} and \bibinfo{author}{S.~Dietrich},
  \enquote{\bibinfo{title}{{Line tension between fluid phases and a
  substrate}}}, \bibinfo{journal}{{Phys. Rev. E}}
  \textbf{\bibinfo{volume}{{57}}}, \bibinfo{pages}{{655--671}}
  (\bibinfo{year}{{1998}}).

\bibitem{BauerDietrich}
\bibinfo{author}{C.~Bauer} and \bibinfo{author}{S.~Dietrich},
  \enquote{\bibinfo{title}{{Quantitative study of laterally inhomogeneous
  wetting films}}}, \bibinfo{journal}{{Eur. Phys. J. B}}
  \textbf{\bibinfo{volume}{{10}}}, \bibinfo{pages}{{767--779}}
  (\bibinfo{year}{{1999}}).

\bibitem{Dobbs/Indekeu}
\bibinfo{author}{H.~Dobbs} and \bibinfo{author}{J.~Indekeu},
  \enquote{\bibinfo{title}{{Line tension at wetting - interface displacement
  model beyond the gradient-squared approximation}}},
  \bibinfo{journal}{{Physica A}} \textbf{\bibinfo{volume}{{201}}},
  \bibinfo{pages}{{457--481}} (\bibinfo{year}{{1993}}).

\bibitem{Churaev}
\bibinfo{author}{N.~Churaev}, \bibinfo{author}{V.~Starov}, and
  \bibinfo{author}{B.~Derjaguin}, \enquote{\bibinfo{title}{{The shape of the
  transition zone between a thin-film and bulk liquid and the line tension}}},
  \bibinfo{journal}{{J. Colloid Interface Sci.}}
  \textbf{\bibinfo{volume}{{89}}}, \bibinfo{pages}{{16--24}}
  (\bibinfo{year}{{1982}}).

\bibitem{DeGennes}
\bibinfo{author}{P.-G. De~Gennes}, \enquote{\bibinfo{title}{{Wetting - Statics
  and Dynamics}}}, \bibinfo{journal}{{Rev. Mod. Phys.}}
  \textbf{\bibinfo{volume}{{57}}}, \bibinfo{pages}{{827--863}}
  (\bibinfo{year}{{1985}}).

\bibitem{Amirfazli}
\bibinfo{author}{A.~Amirfazli} and \bibinfo{author}{A.~W. Neumann},
  \enquote{\bibinfo{title}{Status of the three-phase line tension: a review}},
  \bibinfo{journal}{Advances in Colloid and Interface Science}
  \textbf{\bibinfo{volume}{110}}, \bibinfo{pages}{121 -- 141}
  (\bibinfo{year}{2004}).

\bibitem{Indekeu}
\bibinfo{author}{J.~Indekeu}, \enquote{\bibinfo{title}{{Line Tension near the
  Wetting Transition - Results from an Interface Displacement Model}}},
  \bibinfo{journal}{{Physica A}} \textbf{\bibinfo{volume}{{183}}},
  \bibinfo{pages}{{439--461}} (\bibinfo{year}{{1992}}).

\bibitem{Indekeu1994}
\bibinfo{author}{J.~O. Indekeu}, \enquote{\bibinfo{title}{{Line tension at
  wetting}}}, \bibinfo{journal}{{Int. J. Mod. Phys. B}}
  \textbf{\bibinfo{volume}{{8}}}, \bibinfo{pages}{{309--345}}
  (\bibinfo{year}{{1994}}).

\bibitem{DrelichMiller}
\bibinfo{author}{J.~Drelich} and \bibinfo{author}{J.~Miller},
  \enquote{\bibinfo{title}{{The Effect of Solid Surface Heterogeneity and
  Roughness on the Contact Angle/Drop (Bubble) Size Relationship}}},
  \bibinfo{journal}{{J. Colloid Interface Sci.}}
  \textbf{\bibinfo{volume}{{164}}}, \bibinfo{pages}{{252--259}}
  (\bibinfo{year}{{1994}}).

\bibitem{DrelichMiller92}
\bibinfo{author}{J.~Drelich} and \bibinfo{author}{J.~Miller},
  \enquote{\bibinfo{title}{{The line/pseudo-line tension in three-phase
  systems}}}, \bibinfo{journal}{{Particul. Sci. Technol.}}
  \textbf{\bibinfo{volume}{{10}}}, \bibinfo{pages}{{1--20}}
  (\bibinfo{year}{{1992}}).

\bibitem{Gaydos}
\bibinfo{author}{J.~Gaydos} and \bibinfo{author}{A.~W. Neumann},
  \enquote{\bibinfo{title}{{The Dependence of Contact Angles on Drop Size and
  Line Tension}}}, \bibinfo{journal}{{J Colloid Interface Sci.}}
  \textbf{\bibinfo{volume}{{120}}}, \bibinfo{pages}{{76--86}}
  (\bibinfo{year}{{1987}}).

\bibitem{Li1990}
\bibinfo{author}{D.~Li} and \bibinfo{author}{A.~Neumann},
  \enquote{\bibinfo{title}{{Determination of line tension from the drop size
  dependence of contact angles}}}, \bibinfo{journal}{{Colloids Surfaces}}
  \textbf{\bibinfo{volume}{{43}}}, \bibinfo{pages}{{195--206}}
  (\bibinfo{year}{{1990}}).

\bibitem{Vera-Graziano}
\bibinfo{author}{R.~Vera-Graziano}, \bibinfo{author}{S.~Muhl}, and
  \bibinfo{author}{F.~Rivera-Torres}, \enquote{\bibinfo{title}{The effect of
  illumination on contact angles of pure water on crystalline silicon}},
  \bibinfo{journal}{J. Colloid Interface Sci.} \textbf{\bibinfo{volume}{170}},
  \bibinfo{pages}{591} (\bibinfo{year}{1995}).

\bibitem{Drelich}
\bibinfo{author}{J.~Drelich}, \enquote{\bibinfo{title}{{The significance and
  magnitude of the line tension in three-phase (solid-liquid-fluid) systems}}},
  \bibinfo{journal}{{Colloid Surf. A-Physicochem. Eng. Asp.}}
  \textbf{\bibinfo{volume}{{116}}}, \bibinfo{pages}{{43--54}}
  (\bibinfo{year}{{1996}}).

\bibitem{Checco}
\bibinfo{author}{A.~Checco}, \bibinfo{author}{P.~Guenoun}, and
  \bibinfo{author}{J.~Daillant}, \enquote{\bibinfo{title}{Nonlinear dependence
  of the contact angle of nanodroplets on contact line curvature}},
  \bibinfo{journal}{Phys. Rev. Lett.} \textbf{\bibinfo{volume}{91}},
  \bibinfo{pages}{186101} (\bibinfo{year}{2003}).

\bibitem{Li}
\bibinfo{author}{D.~Li}, \bibinfo{author}{F.~Lin}, and
  \bibinfo{author}{A.~Neumann}, \enquote{\bibinfo{title}{{Effect of
  corrugations of the 3-phase line on the drop size dependence of contact
  angles}}}, \bibinfo{journal}{{J. Colloid Interface Sci.}}
  \textbf{\bibinfo{volume}{{142}}}, \bibinfo{pages}{{224--231}}
  (\bibinfo{year}{{1991}}).

\bibitem{Borkent}
\bibinfo{author}{B.~M. Borkent}, \bibinfo{author}{S.~de~Beer},
  \bibinfo{author}{F.~Mugele}, and \bibinfo{author}{D.~Lohse},
  \enquote{\bibinfo{title}{{On the Shape of Surface Nanobubbles}}},
  \bibinfo{journal}{{Langmuir}} \textbf{\bibinfo{volume}{{26}}},
  \bibinfo{pages}{{260--268}} (\bibinfo{year}{{2010}}).

\bibitem{Yang}
\bibinfo{author}{J.~Yang}, \bibinfo{author}{J.~Duan},
  \bibinfo{author}{D.~Fornasiero}, and \bibinfo{author}{J.~Ralston},
  \enquote{\bibinfo{title}{Very small bubble formation at the solid‚àíwater
  interface}}, \bibinfo{journal}{J. Phys. Chem. B}
  \textbf{\bibinfo{volume}{107}}, \bibinfo{pages}{6139--6147}
  (\bibinfo{year}{2003}).

\bibitem{Kameda}
\bibinfo{author}{N.~Kameda}, \bibinfo{author}{N.~Sogoshi}, and
  \bibinfo{author}{S.~Nakabayashi}, \enquote{\bibinfo{title}{Nitrogen
  nanobubbles and butane nanodroplets at si(1‚Ä†0‚Ä†0)}},
  \bibinfo{journal}{Surface Science} \textbf{\bibinfo{volume}{602}},
  \bibinfo{pages}{1579 -- 1584} (\bibinfo{year}{2008}).

\bibitem{Nakabayashi}
\bibinfo{author}{N.~Kameda} and \bibinfo{author}{S.~Nakabayashi},
  \enquote{\bibinfo{title}{Size-induced sign inversion of line tension in
  nanobubbles at a solid/liquid interface}}, \bibinfo{journal}{Chem. Phys.
  Lett.} \textbf{\bibinfo{volume}{461}}, \bibinfo{pages}{122 -- 126}
  (\bibinfo{year}{2008}).

\bibitem{Pompe}
\bibinfo{author}{T.~Pompe} and \bibinfo{author}{S.~Herminghaus},
  \enquote{\bibinfo{title}{{Three-phase contact line energetics from nanoscale
  liquid surface topographies}}}, \bibinfo{journal}{{Phys. Rev. Lett.}}
  \textbf{\bibinfo{volume}{{85}}}, \bibinfo{pages}{{1930--1933}}
  (\bibinfo{year}{{2000}}).

\bibitem{Zorin}
\bibinfo{author}{Z.~Zorin}, \bibinfo{author}{D.~Platikanov}, and
  \bibinfo{author}{T.~Kolarov}, \enquote{\bibinfo{title}{{The transition region
  between aqueous wetting films on quartz and the adjacent meniscus}}},
  \bibinfo{journal}{{Colloids Surfaces}} \textbf{\bibinfo{volume}{{22}}},
  \bibinfo{pages}{{147--159}} (\bibinfo{year}{{1987}}).

\bibitem{Halverson}
\bibinfo{author}{J.~D. Halverson}, \bibinfo{author}{C.~Maldarelli},
  \bibinfo{author}{A.~Couzis}, and \bibinfo{author}{J.~Koplik},
  \enquote{\bibinfo{title}{{Atomistic simulations of the wetting behavior of
  nanodroplets of water on homogeneous and phase separated self-assembled
  monolayers}}}, \bibinfo{journal}{{Soft Matter}}
  \textbf{\bibinfo{volume}{{6}}}, \bibinfo{pages}{{1297--1307}}
  (\bibinfo{year}{{2010}}).

\bibitem{Ingebrigtsen}
\bibinfo{author}{T.~Ingebrigtsen} and \bibinfo{author}{S.~Toxvaerd},
  \enquote{\bibinfo{title}{Contact angles of lennard-jones liquids and droplets
  on planar surfaces}}, \bibinfo{journal}{J. Phys. Chem. C}
  \textbf{\bibinfo{volume}{111}}, \bibinfo{pages}{8518--8523}
  (\bibinfo{year}{2007}).

\bibitem{gromacs}
\bibinfo{author}{D.~Van~der Spoel}, \bibinfo{author}{E.~Lindahl},
  \bibinfo{author}{B.~Hess}, \bibinfo{author}{G.~Groenhof},
  \bibinfo{author}{A.~Mark}, and \bibinfo{author}{H.~Berendsen},
  \enquote{\bibinfo{title}{{GROMACS: Fast, flexible, and free}}},
  \bibinfo{journal}{{J. Comput. Chem.}} \textbf{\bibinfo{volume}{{26}}},
  \bibinfo{pages}{{1701--1718}} (\bibinfo{year}{{2005}}).

\bibitem{Barrat}
\bibinfo{author}{J.~Barrat} and \bibinfo{author}{L.~Bocquet},
  \enquote{\bibinfo{title}{{Large slip effect at a nonwetting fluid-solid
  interface}}}, \bibinfo{journal}{{Phys. Rev. Lett.}}
  \textbf{\bibinfo{volume}{{82}}}, \bibinfo{pages}{{4671--4674}}
  (\bibinfo{year}{{1999}}).

\bibitem{Dammer}
\bibinfo{author}{S.~M. Dammer} and \bibinfo{author}{D.~Lohse},
  \enquote{\bibinfo{title}{Gas enrichment at liquid-wall interfaces}},
  \bibinfo{journal}{Phys. Rev. Lett.} \textbf{\bibinfo{volume}{96}},
  \bibinfo{pages}{206101} (\bibinfo{year}{2006}).

\bibitem{Nijmeijer}
\bibinfo{author}{M.~Nijmeijer}, \bibinfo{author}{C.~Bruin},
  \bibinfo{author}{A.~Bakker}, and \bibinfo{author}{J.~Van~Leeuwen},
  \enquote{\bibinfo{title}{{Wetting and drying of an inert wall by a fluid in a
  molecular-dynamics simulation}}}, \bibinfo{journal}{{Phys. Rev. A}}
  \textbf{\bibinfo{volume}{{42}}}, \bibinfo{pages}{{6052--6059}}
  (\bibinfo{year}{{1990}}).

\bibitem{Rowlinson/Widom}
\bibinfo{author}{J.~S. Rowlinson} and \bibinfo{author}{B.~Widom},
  \emph{\bibinfo{title}{Molecular Theory of Capillarity}}
  (\bibinfo{publisher}{Clarendon, Oxford}) (\bibinfo{year}{1982}).

\bibitem{Hansen}
\bibinfo{author}{J.-P. Hansen} and \bibinfo{author}{I.~R. McDonald},
  \emph{\bibinfo{title}{Theory of Simple Liquids}}
  (\bibinfo{publisher}{Academic Press, London}) (\bibinfo{year}{2006}).

\bibitem{MerchantKeller}
\bibinfo{author}{G.~J. Merchant} and \bibinfo{author}{J.~B. Keller},
  \enquote{\bibinfo{title}{{Contact Angles}}}, \bibinfo{journal}{{Phys. Fluids
  A}} \textbf{\bibinfo{volume}{{4}}}, \bibinfo{pages}{{477--485}}
  (\bibinfo{year}{{1992}}).

\bibitem{Snoeijer}
\bibinfo{author}{J.~H. Snoeijer} and \bibinfo{author}{B.~Andreotti},
  \enquote{\bibinfo{title}{{A microscopic view on contact angle selection}}},
  \bibinfo{journal}{{Phys. Fluids}} \textbf{\bibinfo{volume}{{20}}},
  \bibinfo{pages}{{057101}} (\bibinfo{year}{{2008}}).

\bibitem{Marmur}
\bibinfo{author}{A.~Marmur}, \enquote{\bibinfo{title}{{Line tension and the
  intrinsic contact angle in solid-liquid-fluid systems}}},
  \bibinfo{journal}{{J. Colloid Interface Sci.}}
  \textbf{\bibinfo{volume}{{186}}}, \bibinfo{pages}{{462--466}}
  (\bibinfo{year}{{1997}}).

\bibitem{BorkentPRL}
\bibinfo{author}{B.~M. Borkent}, \bibinfo{author}{S.~M. Dammer},
  \bibinfo{author}{H.~Sch\"onherr}, \bibinfo{author}{G.~J. Vancso}, and
  \bibinfo{author}{D.~Lohse}, \enquote{\bibinfo{title}{Superstability of
  surface nanobubbles}}, \bibinfo{journal}{Phys. Rev. Lett.}
  \textbf{\bibinfo{volume}{98}}, \bibinfo{pages}{204502}
  (\bibinfo{year}{2007}).

\bibitem{Hampton}
\bibinfo{author}{M.~A. Hampton} and \bibinfo{author}{A.~V. Nguyen},
  \enquote{\bibinfo{title}{{Nanobubbles and the nanobubble bridging capillary
  force}}}, \bibinfo{journal}{{Adv. Colloid Interface Sci.}}
  \textbf{\bibinfo{volume}{{154}}}, \bibinfo{pages}{{30--55}}
  (\bibinfo{year}{{2010}}).

\bibitem{Brenner}
\bibinfo{author}{M.~P. Brenner} and \bibinfo{author}{D.~Lohse},
  \enquote{\bibinfo{title}{Dynamic equilibrium mechanism for surface nanobubble
  stabilization}}, \bibinfo{journal}{Phys. Rev. Lett.}
  \textbf{\bibinfo{volume}{101}}, \bibinfo{pages}{214505}
  (\bibinfo{year}{2008}).

\end{thebibliography}

\end{document}